# 4G/5G CELL-LEVEL MULTI-INDICATOR FORECASTING BASED ON DENSE-MLP

Jiacheng Yin[1], Wenwen Li[1], Xidong Wang[1], Xiaozhou Ye[1], Ye Ouyang[1]
[1]Asiainfo Technologies (China), Inc. Beijing, China

NOTE: Corresponding author: Wenwen Li, liww7@asiainfo.com

**Abstract** – With the development of 4G/5G, the rapid growth of traffic has caused a large number of cell indicators to exceed the warning threshold, and network quality has deteriorated. It is necessary for operators to solve the congestion in advance and effectively to guarantee the quality of user experience. Cell-level multi-indicator forecasting is the foundation task for proactive complex network optimization. In this paper, we propose the 4G/5G Cell-level multi-indicator forecasting method based on the dense-Multi-Layer Perceptron (MLP) neural network, which adds additional fully-connected layers between non-adjacent layers in an MLP network. The model forecasted the following week's traffic indicators of 13000 cells according to the six-month historical indicators of 65000 cells in the 4G&5G network，which got the highest weighted MAPE score (0.2484) in the China Mobile problem statement in the ITU-T AI/ML in 5G Challenge 2021. Furthermore, the proposed model has been integrated into the AsiaInfo 4G/5G energy-saving system and deployed in Jiangsu Province of China.

## 1. INTRODUCTION

With the development of 4G/5G, the rapid growth of traffic has caused a large number of cell indicators to exceed the warning threshold, and the network quality has deteriorated. The daily operations for network optimization mainly consider the following key points: (1) forecast network usage to avoid or reduce the probability of network congestion, improve network resource allocation efficiency, and ensure high-quality user experience; (2) evaluate the quality of 4G/5G networks; (3) reduce the load of high-capacity cells (mainly 4G at this stage) to achieve active optimization. Among them, cell-level multi-indicator forecasting is the foundational task for proactive complex network optimization.

In view of the high investment cost of network infrastructure, the network capacity cannot be infinitely enlarged. In order to avoid network overload, the network capacity should be partially controlled. It has become the most economical and effective method to forecast and evaluate network quality and capacity by using a big data forecasting method for a large number of network performance indicator data collected in different cities and different systems to identify areas with insufficient network performance and carry out targeted optimization or concentrated investment and construction.

ITU has specified some network intelligence use cases of cell traffic forecasting. ITU-T M.3080 [4] provides a framework for artificial intelligence enhanced telecommunication operation and management (AITOM). ITU-T Y.3172 [5] specifies an architectural framework for Machine Learning (ML) in future networks including IMT-2020. ITU-T Y.3173 [6] specifies a framework for evaluating the intelligence of future networks including IMT-2020. ITU-T Y.3175 [7] specifies a functional architecture of Quality of Service (QoS) assurance based on Machine Learning (ML) for the International Mobile Telecommunications-2020 (IMT-2020) network.

At present, for 4G and 5G cells, six traffic KPIs [9] are mainly used for cell capacity identification: Physical Uplink Shared Channel (PUSCH), Physical Downlink Shared Channel (PDSCH), Physical Downlink Control Channel (PDCCH), average of valid Radio Resource Control (RRC) connections , Packet Data Convergence Protocol Uplink Flow (PDCPUL) and Packet Data Convergence Protocol Downlink Flow (PDCPDL).

By mining the correlation between the historical data of 4G/5G cell-level indicators in different regions and cities, we aim to construct a cell-level multi-indicator forecasting model to forecast the trend of the indicators with respect to each cell.

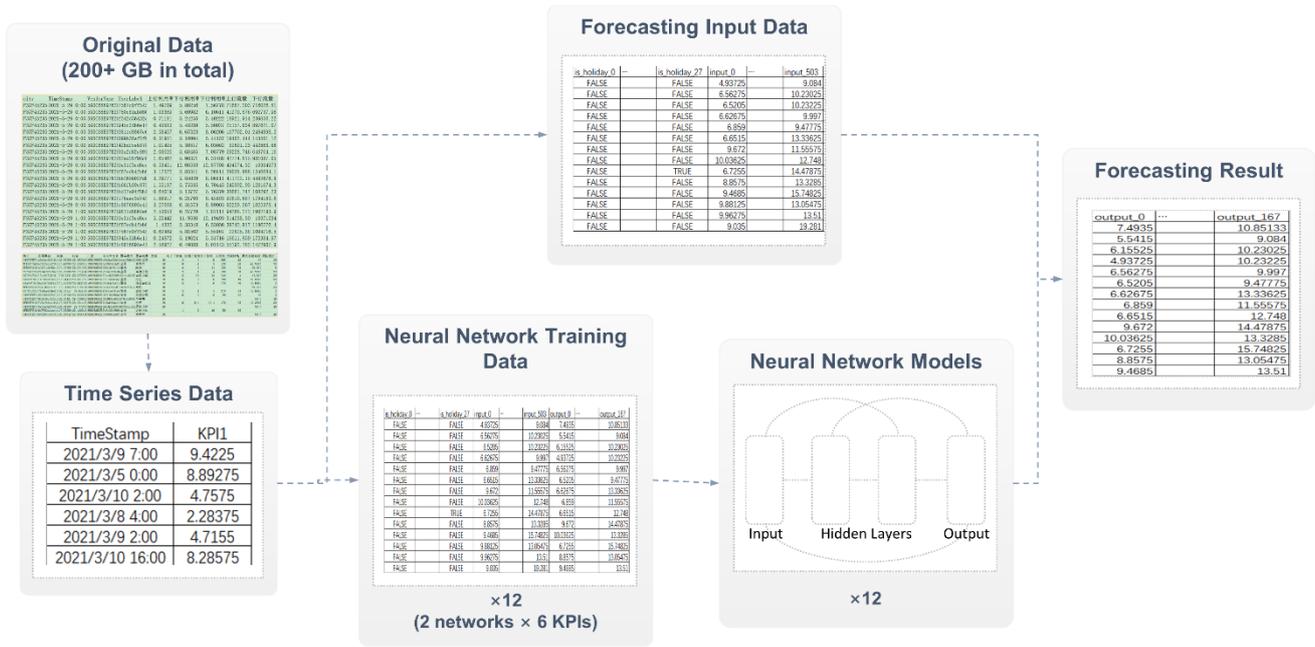

**Fig. 1** – Solution overview

After prediction, if any cell reaches the indicator early warning threshold, the cell needs to be shunted to reduce cell load and improve network quality.

## 2. RELATED WORK

### 2.1 Time series forecasting methods

Researchers have proposed many classical time series forecasting algorithms for such cases

Prophet [1] is a procedure for forecasting time series data based on an additive model where nonlinear trends are fit with yearly, weekly, and daily seasonality, plus holiday effects.

LSTNet [3] uses a Convolution Neural Network (CNN) and Recurrent Neural Network (RNN) to extract short-term local dependency patterns among variables and discover long-term patterns for time series trends.

Informer [2] is an efficient transformer-based model for Long Sequence Time-series Forecasting (LSTF), with three distinctive characteristics: (1) a ProbSparse self-attention mechanism; (2) the self-attention distilling highlights; (3) the conceptually simple generative style decoder. Informer forecasts long time-series sequences at one forward operation rather than in a step-by-step way, which drastically improves the inference speed of long-sequence forecastings.

DeepAR [12] is a methodology for producing accurate probabilistic forecasts, based on training an auto-regressive recurrent network model on a large number of related time series. It effectively learns a global model from a related time series, handles widely-varying scales through rescaling and velocity-based sampling, generates calibrated probabilistic forecasts with high accuracy, and can learn complex patterns such as seasonality and uncertainty growth over time from the data.

### 2.2 Researches that apply "dense" skip-connections on neural networks

The exploration of network architectures has been a part of neural network research since their initial discovery. As neural networks become increasingly deep, a new problem emerges: as information about the input or gradient passes through many layers, it can vanish and "wash out" by the time it reaches the end (or beginning) of the network. DenseNet[8] connects all layers directly with each other. To preserve the feedforward nature, each layer obtains additional inputs from all preceding layers. This architecture distills the insight into a simple connectivity pattern: to ensure maximum information flow between layers in the networks.

DenseNet is a CNN-based network design used for image classification problems, and some recent researches have attempted to apply DenseNet-style skip connections on non-convolutional neural networks and use them in real-world applications.

AdnFM [13] combines DenseNet-style residual learning and an attention mechanism to predict Click-Through-Rate (CTR). Steinholtz [14] applied skip connections on MLP models for classifications of Parkinson's disease.

The idea of adding such skip connections to neural networks has also been experimented in an earlier work in which Raiko, Valpola, and LeCun added linear normalization terms to a tanh activation function together with the skip connections in neural network models for multiple tasks [15].

## 3. SOLUTIONS

Our solution is based on a custom neural network called "Dense-MLP" consisting of data preprocessing, feature engineering, model training, and forecasting as shown in Fig. 1.

In the data preprocessing part, we convert original data into time series data and fill the missing values according to the time series' periodicities. In the feature engineering part, we convert the time series into tabular data that can be used to train a regression model. During the model training, we train our dense-MLP model with the tabular data, and during forecasting, we use the trained models to forecast the final results.

### 3.1 Data preprocessing

#### 3.1.1 Data set

Our methods are tested on a machine learning competition [9], whose training data set includes hourly 4G/5G cell indicators data of four cities in a province of China from Jan 1 2021, to Jun 31 2021. The detailed information is shown in Table 1.

**Table 1** – Detailed information for the data set

| pattern | #cell | | | | #indicators | data size/GB |
|---|---|---|---|---|---|---|
| | city1 | city2 | city3 | city4 | | |
| 4G | 24374 | 22932 | 10796 | 5223 | 6 | 191+ |
| 5G | 1714 | 1787 | 652 | 207 | 6 | 25+ |

#### 3.1.2 Missing values

Based on the experience of domain experts and observation of the data, we found that all the six indicators that we were trying to forecast (PUSCH, PDSCH, PDCCH, RRC, PDCPUL, and PDCPDL) contain obvious daily and weekly periodicities, as shown in Fig. 2.

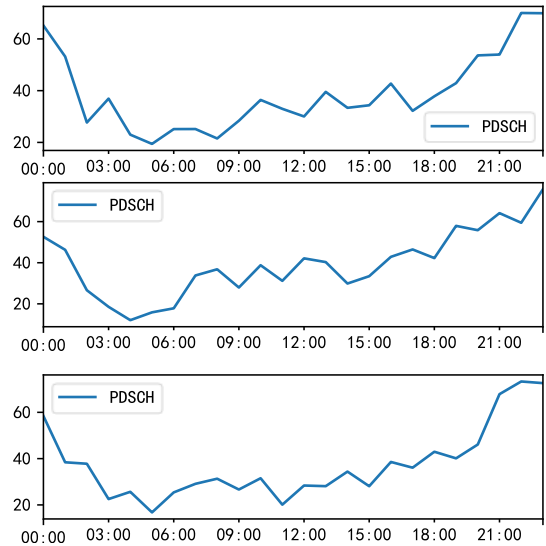

(a) Daily periodicity

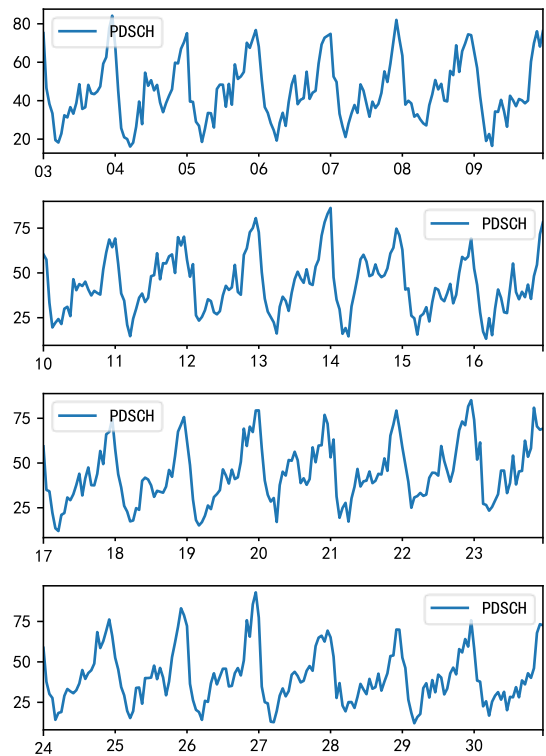

(b) Weekly periodicity

**Fig. 2** – Example of periodicity implied in the indicator

According to these periodicities, we used a weighted average of the values at the same time point of each week before and after the missing value as the filling value, with the weights inversely proportional to the number of weeks that differ from the time point of the missing value. For example, for hourly data from March 1 to March 31, 2021, if the data at 05:00 on March 10 is missing, the weighted average of the data at 05:00 on March 3, 05:00 on March 17, 05:00 on March 24, and 05:00 on March 31 is taken as the filling value, with their un-normalized weights being 1, 1, 1/2, 1/3, as shown in Fig. 3.

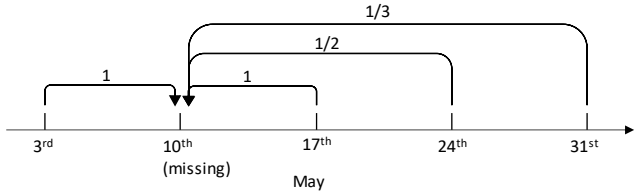

**Fig. 3** – Method for filling missing values

### 3.1.3 Feature engineering

The dense-MLP model that we chose is a neural-network-based regression model, whose input and output features need to be extracted from the time series before training.

The model tries to forecast values for the following week from three weeks' historical data of a single traffic KPI. The input of the model is the holiday features of all the four weeks (1 means the day is a holiday and 0 means it's not) and the traffic KPI values of the first three weeks, including 28 + 24*21 = 532 input features in total, as shown in Table 2. The model's output is the KPI values of the next week, including 24*7=168 output features, as shown in Table 3.

**Table 2** – One row of input features

| Holiday feature 1 | ... | Holiday feature 28 | Historical value 1 | ... | Historical value 504 |
|---|---|---|---|---|---|
| 1 | ... | 0 | 30.503 | ... | 25.210 |

**Table 3** – One row of output features

| Forecast 1 | ... | Forecast 168 |
|---|---|---|
| 10.133 | ... | 65.989 |

From each traffic KPI's time series, we could extract multiple rows of training data in a sliding-window manner, with each row containing all 532+168=700 features. For example, if we need to train models on data extracted from 10,000 bases from March 1 to March 31 2021, then four sample rows can be extracted from March 1 2021, to March 28 2021, March 2 2021 to March 29 2021, March 3 2021, to March 30 2021, and March 4 2021 to March 31 2021, accordingly. Thus, a total of 10000*4=40000 rows of training data can be obtained.

During training, input and output features of the model need to be scaled: for each row of input features, the 504 historical values in the input features are divided by the average of the 504 features as the scaled input features, and the model outputs are multiplied by the average value as the final forecastings.

### 3.2 Dense-MLP neural network model

The dense-MLP model described in this section was intuitively inspired by DenseNet [8], which includes "dense" skip connections between network modules. Based on the MLP feedforward neural network, our model includes additional fully connected layers added between non-adjacent layers:

$$\boldsymbol{h_1} = ReLU(\boldsymbol{I} \cdot \boldsymbol{W_1}) \quad (1)$$

$$\boldsymbol{h_2} = ReLU\,(\boldsymbol{I} \cdot \boldsymbol{W_2} + \boldsymbol{h_1} \cdot \boldsymbol{W_3})/2 \quad (2)$$

$$\boldsymbol{O} = (\boldsymbol{I} \cdot \boldsymbol{W_4} + \boldsymbol{h_1} \cdot \boldsymbol{W_5} + \boldsymbol{h_2} \cdot \boldsymbol{W_6})/3 \quad (3)$$

Where $\boldsymbol{I}$ is the scaled model input vector, $\boldsymbol{h_1}$ and $\boldsymbol{h_2}$ are the outputs of each hidden layer, $\boldsymbol{O}$ is the scaled model output vector, and $\boldsymbol{W_1}$-$\boldsymbol{W_6}$ are the parameter matrices of each fully connected layer. The neural network archietecture is shown in Fig. 4.

The hidden layers use a ReLU activation function, with the number of neurons in each hidden layer being 4096.

During reasoning, taking into consideration that the values of the six indicators that we were trying to forecast should all be non-negative, we also applied a ReLU function on the output layer.

Intuitively, the dense-MLP network preserves the linear relationship between inputs and outputs better than the MLP model, thus reducing the information loss caused by the feeding forward process through multiple FC layers. Compared with other neural network architectures proposed for time series forecasting of single or a few time series, the dense-MLP model contains a larger number of trainable parameters and is not limited by the locality assumption of convolutional neural networks and the long-term memory capabilities of recursive neural network structures (RNN [10], LSTM [11], etc.), allowing it to better fit the nonlinear long-term dependencies within the massive KPI data from the telecommunications industry.

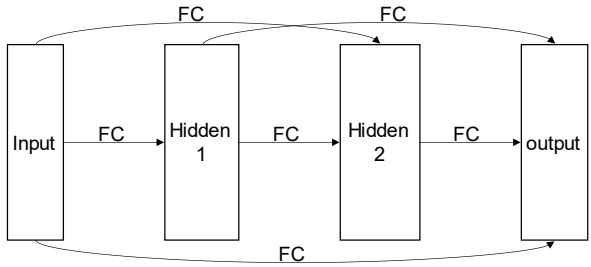

**Fig. 4** – The dense-MLP model structure

### 3.3 Training loss

Since the telecommunication operators are more concerned with cases when the forecasting results exceed the warning thresholds, a weighted MAPE score was chosen as the evaluation matrix of the competition (See Subsection 4.1). However, considering that MAPE is rather sensitive to the errors when the true value is small, our solution uses a customized loss function during training, in which MAPE is replaced by a "unified absolute error" when the Absolute Percentage Error (APE) exceeds 100%, which produces a relatively limited gradient where the true value is small so that the model converges on all traffic KPIs. The "unified absolute error" is calculated by dividing the absolute error by the mean of true values in a row instead of individual true values.

$$AE = |y_{actual} - y_{pred}| \quad (4)$$

$$APE = \frac{AE}{|y_{actual}|} \times 100 \quad (5)$$

$$UAE = \frac{AE}{mean(y_{actual})} \times 100 \quad (6)$$

$$CE = \begin{cases} APE, & APE \leq 100 \\ UAE, & APE > 100 \end{cases} \quad (7)$$

$$Loss\ Func = (mean(CE))^2 \quad (8)$$

In the formulas above, $y_{actual}$ means the true values, $y_{pred}$ means the forecasted values, **AE** means absolute error, **APE** means absolute percentage error, **UAE** means the "unified absolute error", **CE** means a combined error function, **Loss Func** means the final loss function used to train the models.

### 3.4 Training hyper-parameters

We used the Adam optimizer with an initial learning rate of 0.005, and the learning rate decreases linearly to close to 0 in 128 epochs of training; that is, the learning rate of the first epoch was 0.0005, and the learning rate of each subsequent epoch was 0.0005/128 = 0.00000390625 smaller than that of the previous epoch.

We set batch size to 8192 or 16384 for different indicators, choosing the one that produces a smaller train set loss after the 128 epochs.

## 4. RESULTS

The method described in Section 3 has been chosen for and evaluated by a machine learning competition [9] which provides time-series data for training (details of the training set described in Subsection 3.1.1) and a private validation data set for evaluation. Our method ranked 1st in the competition, which is our main motivation for publishing the method.

Subsection 4.2 describes the metric used by the competition for evaluating the results, Subsection 4.3 describes our earlier models submitted to the competition, Subsection 4.4 shows the results produced by methods in Subsection 4.3, and Subsection 4.5 describes visualizations of the method.

### 4.1 Metric

The forecasting task involves the forecasting of 63,329 cells for each 4G indicator and 4,364 cells for each 5G cell indicator. Taking business needs into consideration, the weight of 4G cell indicators was $A_{4G}$=0.7 and that for 5G indicators is $A_{5G}$=0.3. The weight on the first day of the 7 days to be forecasted was 1.2 and the weight of the last 6 days was 1.0.

The weighted MAPE error is the weighted average of 4G cell MAPE errors and 5G cell MAPE errors.

$$weighted\_MAPE = A_{4G} \cdot MAPE_{4G} + A_{5G} \cdot MAPE_{5G} \quad (9)$$

### 4.2 Training

In the data preprocessing stage, we used the preprocessing method described in Subsection 3.1 to extract 30 rows of data with a length of 28 days from the data of each cell from March 1 to June 31 for training. About 1.9 million rows of training data can be obtained for each 4G indicator, and about 130,000 rows of training data can be obtained for each 5G indicator. Since the preprocessed data extracted from 5G indicators are smaller in amount, their models are incrementally trained based on the corresponding model for the same 4G cell indicator.

We applied the training hyper-parameters described in Subsection 3.4, and the batch size that minimizes the training loss was used for each indicator.

## 4.3 Models for comparison

Naive model: Simply use the historical data of the past week as the forecasted value for the next week; that is, use the data of June 24 to June 30 as the forecasted value for July 1 to July 7.

Rule-based forecasting model: Based on the periodicities that we observed (daily and weekly), use historical data and apply different statistical operators to carry out a weighted average as the final forecast value. See Appendix I for more details.

DenseMLP-final: Our final solution described in this article.

DenseMLP-history1: Compared to DenseMLP-Final, DenseMLP-history1 has the number of hidden layer neurons reduced to 2048 and other hyper-parameters unchanged.

DenseMLP-history2: Compared to DenseMLP- Final, DenseMLP-history2 has the number of hidden layer neurons reduced to 1024, batch size fixed to 8192, and the initial learning rate increased to 0.001 and other hyper-parameters unchanged.

DenseMLP-history3: Compared to DenseMLP-history2, DenseMLP-history3 has the learning rate fixed to 0.001 instead of linearly decreasing during training and other hyper-parameters unchanged.

## 4.4 Results

Table 4 – Comparing results of our final and earlier methods

| Model | Weighted MAPE |
|---|---|
| Naive | 0.5516 |
| Rule-based | 0.2795 |
| DenseMLP-history3 | 0.2631 |
| DenseMLP-history2 | 0.2521 |
| DenseMLP-history1 | 0.2491 |
| DenseMLP-final | 0.2484 |

By comparing DenseMLP -history1 and DenseMLP – final in Table 4, it can be seen that the DenseMLP models' performance does not strongly depend on the number of hidden layer neurons. In practice, the number of model parameters can be greatly reduced by reducing the hidden layer sizes.

By comparing DenseMLP-history3 and DenseMLP-history2, it can be seen that the addition of a dynamic learning rate significantly improves the result.

The rule-based method achieves a relatively good result, which indicates that our assumption of data periodicity (data includes obvious daily periodicity and certain weekly periodicity) is correct.

Training all 12 DenseMLP-final models on preprocessed data described in Subsection 4.2 takes about 4 hours, and reasoning on preprocessed data which produces the 7-day forecasting described in Subsection 4.1 takes about 2 minutes.

Table 5 – Leaderboard of the Competition [16]

| Method | Weighted MAPE |
|---|---|
| Our method (DenseMLP-final) | 0.2484 |
| 2nd team's method | 0.2560 |
| 3rd team's method | 0.2579 |
| 4th team's method | 0.2597 |
| 5th team's method | 0.2634 |

The leaderboard (Table 5) shows that all leading methods' weighted MAPE scores are near 0.25, which indicates that such errors might be inevitable in this particular data set. Our method outperforms the second team by 0.0076, which looks relatively significant compared to the difference between the following teams (0.0019 between 2nd and 3rd team, 0.0018 between 3rd and 4th team, etc.). All leading teams' methods perform better than our simpler methods ("Naive" or "Rule-based").

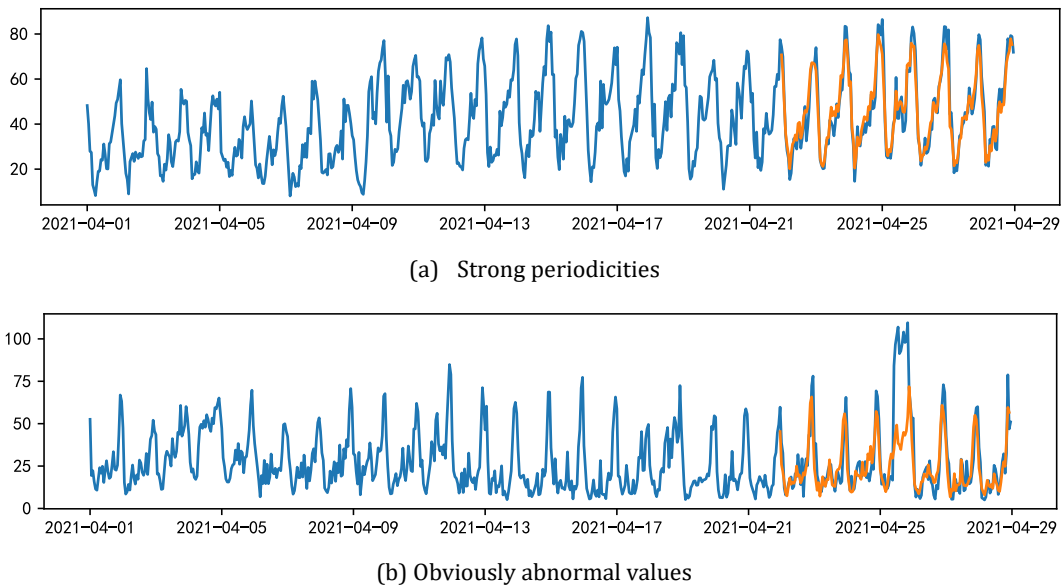

(a) Strong periodicities

(b) Obviously abnormal values

**Fig. 5** – A visualization of the forecasting result of trained dense-MLP models on the training set, in which blue lines are actual values and orange lines are forecasted values.

## 4.5 Visualization

Visualizations in Fig. 5 are examples that show that the models produce reasonable and precise forecasting on "normal" data with strong periodicities, while it doesn't fit obviously abnormal values in the training set.

## 5. APPLICATION AND CONCLUSION

The 4G/5G cell traffic KPIs forecasting model has been integrated into the AsiaInfo energy-saving system and deployed in Jiangsu Province. As shown in Fig. 6, the model can forecast the service quality indicators, traffic indicators, and coverage effect in the energy-saving analysis stage. The accurately and timely forecasting could improve energy-saving benefits and avoid adverse effects on the network.

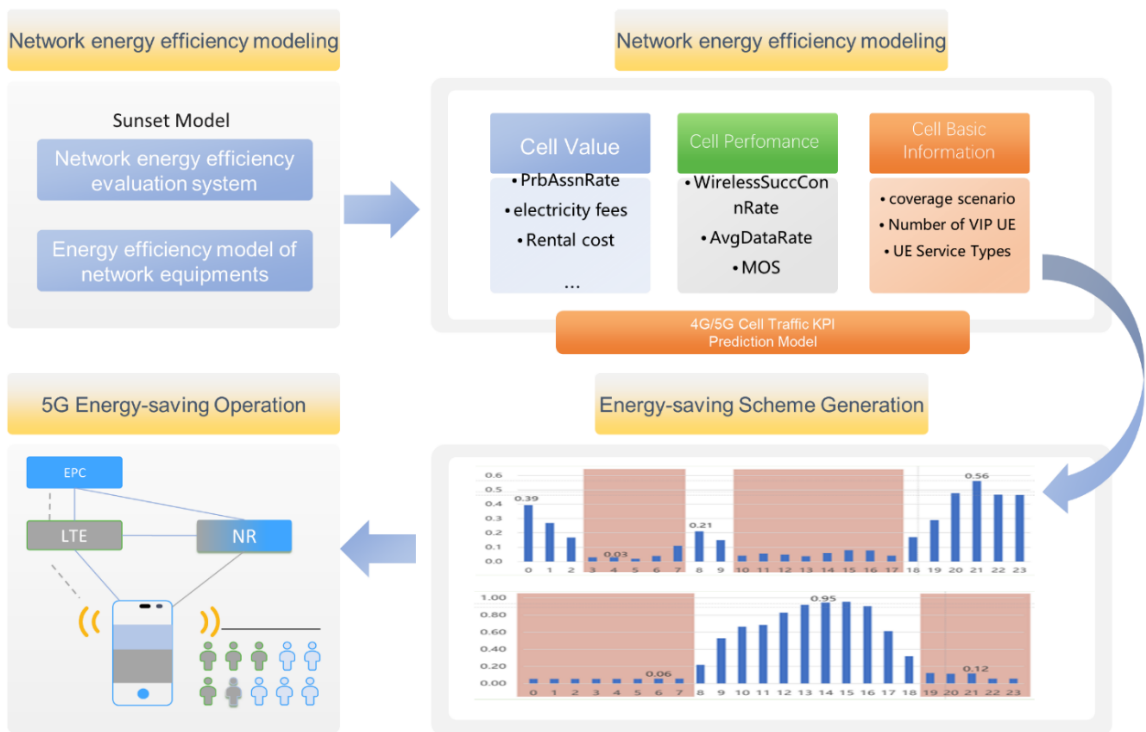

**Fig. 6** – Application scenarios

In future studies and applications, we will experiment with other options on the solution's overall design, model structures, and hyper-parameters. parameters, such as spatial-temporal forecasting which takes the bases' spatial location into consideration, and other deep learning model architectures such as CNN, transformer, and causal convolution. In order to build autonomous networks and improve network intelligence levels, the 4G/5G cell traffic KPIs forecasting model will be widely applied to intelligent 5G network operation and maintenance products.

## Appendix I. Details of our rule-based method

Define an exponential smoothing function:

$$exp\_smooth(x, \alpha) = \sum_{i=1}^{n-1}(1-\alpha)^{n-i}x_i + \alpha x_n \quad (A.1)$$

In the above formula, $x$ is a time series consisting of $(t_1, x_1), \ldots, (t_n, x_n)$ in which $t_i$ is the timestamp at index i and $x_i$ is the value at index i, and $\alpha$ is the smoothing parameter.

Also, we denote the standard mean and median function applied on $x$'s values as $mean(x)$ and $median(x)$.

Given a training time series $S = \{(t_i, s_i)\}$, our rule-based method predicts the value $s_j$ at timestamp $t_j$ in the following way:

$$S_1 = \{s_i | time\_in\_day(t_i) = time\_in\_day(t_j)\} \quad (A.2)$$

$$S_2 = \{s_i | time\_in\_week(t_i) = time\_in\_week(t_j)\} \quad (A.3)$$

$$s_j = w_1 \cdot exp\_smooth(S_1, \alpha_1) + w_2 \cdot exp\_smooth(S_2, \alpha_2) + w_3 \cdot mean(S_1) + w_4 \cdot mean(S_2) + w_5 \cdot median(S_1) + w_6 \cdot median(S_2)$$
$$(A.4)$$

$S_1$ is the sub-series of $S$ consisting of the values at the same time in each day as $t_j$. $S_2$ is the sub-series of $S$ consisting of the values at the same time in each week as $t_j$. $w_1, w_2, \ldots, w_6$ and $\alpha_1$ and $\alpha_2$ are the adjustable parameters of the method.

In the experiments described in Section 4, we set $\alpha_1 = \alpha_2 = 0.82$, $w_1 = 0.07$, $w_2 = 0.13$, $w_3 = 0.14$, $w_4 = 0.26$, $w_5 = 0.14$, $w_6 = 0.26$ to produce a "relatively good" result, as described in Subsection 4.4.

## AUTHORS

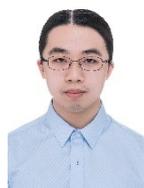
**Jiacheng Yin** M.S. in computing, School of Computer Science, Australian National University, Canberra, Australia; data scientist of Telco AI Lab, AsiaInfo Technologies, with research interests including machine learning, data mining, time series forecasting, etc.

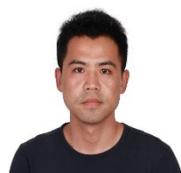
**Wenwen Li** M.S. in control science and engineering, School of Beijing Institute of Technology, Beijing, China; senior data scientist of Telco AI Lab, AsiaInfo Technologies, with research interests including machine learning, data mining, time series forecasting, etc.

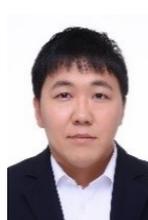
**Xidong Wang** M.S. in communication and information systems, School of Beijing University of Posts and Telecommunications. He is a project manager at Telco AI Lab, AsiaInfo Technologies, His research interests include wireless techniques, 5G communications, machine learning, and its applications.

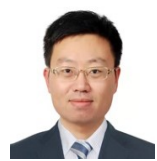
**Xiaozhou Ye** Ph. D., Director of Telco AI Lab, Asiainfo Technology (China) Co., LTD;, his research interests are communication artificial intelligence and future network technology.

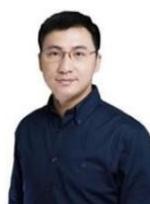
**Ye Ouyang** Ph.D., CTO & Senior Vice President of AsiaInfo Technologies. Dr. Ouyang has distinguished experience in R&D and management in the telecommunications industry. Prior to AsiaInfo, Dr. Ouyang has been a Verizon Fellow and senior manager at Verizon. His research is in the interdisciplinary area of wireless communications, data science, and AI. Dr. Ouyang has authored more than 30 academic papers, 40 patents, 10 international standards, and 8 books.